\documentclass[epj]{webofc}\usepackage[varg]{txfonts}
\usepackage{bm}\usepackage{slashed}
\usepackage{color}\definecolor{violet}{rgb}{0.4336,0,1}
\woctitle{CONF12}

\begin{document}\title{Goldstone-Type Pseudoscalar
Mesons:\\Instantaneous Bethe--Salpeter Models}\author{Wolfgang
Lucha\inst{1}\fnsep\thanks{\email{Wolfgang.Lucha@oeaw.ac.at}}}
\institute{Institute for High Energy Physics, Austrian Academy of
Sciences, Nikolsdorfergasse 18, A-1050 Vienna, Austria}

\abstract{A nontrivial instantaneous variant of the
Bethe--Salpeter formalism allows us to discuss massless
pseudoscalar mesons from a configuration-space-potential
point~of~view.}\maketitle

\section{Incentive: pions and kaons as Goldstonic quark--antiquark
bound states}\label{Sec:I}Quantum field theory describes bound
states of two fermions, of particle and relative momenta $p_{1,2}$
and $p,$ ignoring their total momenta, by Bethe--Salpeter
amplitudes $\Phi(p)$ solving Bethe--Salpeter equations
\begin{equation}\Phi(p)=\frac{\rm i}{(2\pi)^4}\,S_1(p_1)\int{\rm
d}^4q\,K(p,q)\,\Phi(q)\,S_2(-p_2)\label{Eq:BS}\end{equation}
defined by kernels $K(p,q)$ subsuming the underlying interactions,
and fermion propagators $S_{1,2}(p),$ by Lorentz covariance given
by mass and wave-function renormalization functions $M_{1,2}(p^2)$
and $Z_{1,2}(p^2)$:\begin{equation}S_{1,2}(p)=\frac{{\rm
i}\,Z_{1,2}(p^2)}{\slashed{p}-M_{1,2}(p^2)+{\rm i}\,\varepsilon}\
,\qquad\slashed{p}\equiv p^\mu\,\gamma_\mu\ ,\qquad\varepsilon
\downarrow0\ .\label{Eq:P}\end{equation}Potentials from inversion
\cite{WL13} offer an intuitive view at light pseudoscalar mesons
as quark bound states and Goldstone bosons due to spontaneous
chiral symmetry breaking of quantum chromodynamics \cite{WLPoS}.

\section{Instantaneous bound-state equation remembering
Bethe--Salpeter origin}\label{Sec:L}Already some time ago, in an
attempt to retain in the resulting three-dimensional formalism as
much as conceivable of all the information of dynamical origin
encoded in the propagators (\ref{Eq:P}) of the considered
bound-state constituents, we proposed a kind of (relativistically)
least-damage instantaneous reduction of the Bethe--Salpeter
framework \cite{WL05}. Within our approach, we achieve this
objective \emph{not\/} by setting the propagator functions
$M(p^2)$ and $Z(p^2)$ as a whole equal to constants, no longer
aware of any dynamical effects. Rather, we only discard any
dependence of the propagator functions on the time component $p_0$
of the relative momentum $p.$ In other words, when dealing with
$M(p^2)$ and $Z(p^2),$ we place, tentatively, our wagers on some
type of ``$p_0^2=0$ approximation.'' In this case, and under the
equal-time assumption that all interactions incorporated by their
integral kernel $K(p,q)$ are instantaneous --- tantamount to the
requirement that $K(p,q)$ does not depend on the time components
of the relative momenta $p$~and $q,$ i.e.,
$K(p,q)=K(\bm{p},\bm{q})$ ---, the Bethe--Salpeter equation
(\ref{Eq:BS}) can be integrated with respect to $p_0;$ the result
of the integration is a three-dimensional bound-state equation
governing the so-called Salpeter~amplitude\pagebreak
$$\phi(\bm{p})\propto\int{\rm d}p_0\,\Phi(p)\ .$$Expressed by use
of one-fermion free energies $E_{1,2}(p)$ and
positive/negative-energy projectors $\Lambda_{1,2}^\pm(\bm{p}),$
$$E_{1,2}(p)\equiv\sqrt{p^2+M_{1,2}^2(p^2)}\ ,\qquad
\Lambda_{1,2}^\pm(\bm{p})\equiv\frac{E_{1,2}(p)\pm\gamma_0\,
[\bm{\gamma}\cdot\bm{p}+M_{1,2}(p^2)]}{2\,E_{1,2}(p)}\ ,\qquad
p\equiv|\bm{p}|\ ,$$for a bound state of spin-$\frac{1}{2}$
fermion and spin-$\frac{1}{2}$ antifermion, with relative momentum
$\bm{p}$ and bound-state mass $\widehat M,$ our instantaneous
Bethe--Salpeter equation \cite{WL05} is, in the center-of-momentum
frame,~given~by\begin{equation}\phi(\bm{p})=Z_1(p^2)\,Z_2(p^2)
\left(\frac{\Lambda_1^+(\bm{p})\,\gamma_0\,I(\bm{p})\,
\Lambda_2^-(\bm{p})\,\gamma_0}{\widehat M-E_1(p)-E_2(p)}
-\frac{\Lambda_1^-(\bm{p})\,\gamma_0\,I(\bm{p})\,
\Lambda_2^+(\bm{p})\,\gamma_0}{\widehat M+E_1(p)+E_2(p)}\right)\
,\qquad p\equiv|\bm{p}|\ ,\label{Eq:CM}\end{equation}abbreviating,
for ease of notation, the three-dimensional remainder of the
interaction term in Eq.~(\ref{Eq:BS})~by$$I(\bm{p})\equiv
\frac{1}{(2\pi)^3}\int{\rm d}^3q\,K(\bm{p},\bm{q})\,\phi(\bm{q})\
.$$The comparison with predictions obtained upon enabling the
analytic integrability of the instantaneous limit of
Eq.~(\ref{Eq:BS}) with respect to $p_0$ by approximation of both
propagators $S_{1,2}$ by just their free forms~\cite{WL00} shows
that describing bound states of quarks \cite{WLQ} by
Eq.~(\ref{Eq:CM}) has visible impact on their mass
spectra~\cite{WL05a}.

\section{Configuration-space interquark potential: straightforward
determination}\label{Sec:N}In principle, within the framework of
Sec.~\ref{Sec:L} the way how to extract underlying interquark
potentials~by inversion is pretty clear. For simplicity, let's
focus on bound states composed of particle and associated
antiparticle, so that the indices $1,2$ may be dropped. In this
case, our bound-state equation (\ref{Eq:CM}) becomes
\begin{equation}\phi(\bm{p})=Z^2(p^2)\left(
\frac{\Lambda^+(\bm{p})\,\gamma_0\,I(\bm{p})\,\Lambda^-(\bm{p})\,
\gamma_0}{\widehat M-2\,E(p)}-
\frac{\Lambda^-(\bm{p})\,\gamma_0\,I(\bm{p})\,\Lambda^+(\bm{p})\,
\gamma_0}{\widehat M+2\,E(p)}\right)\ ,\qquad p\equiv|\bm{p}|\
.\label{Eq:F}\end{equation}

For each pseudoscalar meson that is a bound state built up by a
quark and corresponding antiquark, its Salpeter amplitude
$\phi(\bm{p})$ can be cast, in terms of two\footnote{Pseudoscalar
mesons are states of zero spin and relative orbital angular
momentum; by inspection of the charge-conjugation and parity
quantum numbers related, in the decomposition over a complete set
of Dirac matrices, to the \emph{independent\/} components of the
Salpeter amplitude, we realize that only two of the latter are
compatible with the expected behaviour of spin-singlet states.}
independent components $\varphi_{1,2}(\bm{p}),$ into the form
$$\phi(\bm{p})=\left[\varphi_1(\bm{p})\,
\frac{\gamma_0\,[\bm{\gamma}\cdot\bm{p}+M(p^2)]}{E(p)}
+\varphi_2(\bm{p})\right]\gamma_5\ ,\qquad
E(p)\equiv\sqrt{p^2+M^2(p^2)}\ ,\qquad p\equiv|\bm{p}|\ .$$The
kernel, $K(\bm{p},\bm{q}),$ comprises Lorentz nature and momentum
dependence of the effective interactions between the bound-state
constituents; assuming identical couplings to quark and antiquark,
in terms of generalized Dirac matrices $\Gamma$ and associated
Lorentz-scalar potentials $V_\Gamma(\bm{p},\bm{q}),$ its action on
$\phi(\bm{q})$~reads$$K(\bm{p},\bm{q})\,\phi(\bm{q})=\sum_\Gamma
V_\Gamma(\bm{p},\bm{q})\,\Gamma\,\phi(\bm{q})\,\Gamma\ .$$We fix
the Dirac nature by use of the
Fierz-symmetry-enforcing\footnote{\emph{Fierz symmetry\/} is the
invariance of Lorentz-scalar quartic products of Dirac field
\emph{operators\/} $\psi_{1,\dots,4}(x)$ under a rearrangement of
$\psi_{1,\dots,4}(x);$ formulated in terms of Dirac matrices
$\Gamma_{i=1,\dots,16},$ $\sum_i\bar\psi_1(x)\,\Gamma_i\,\psi_2(x)
\,\bar\psi_3(x)\,\Gamma_i\,\psi_4(x)=\sum_i\bar\psi_1(x)\,\Gamma_i
\,\psi_4(x)\,\bar\psi_3(x)\,\Gamma_i\,\psi_2(x).$} linear
combination of tensor products$$\Gamma\otimes\Gamma=\frac{1}{2}\,
(\gamma_\mu\otimes\gamma^\mu+\gamma_5\otimes\gamma_5-1\otimes1)\
.$$Assuming the potential $V_\Gamma(\bm{p},\bm{q})$ to be of
convolution type and to exhibit spherical symmetry enables us to
split off all dependence on angular variables.
Equation~(\ref{Eq:F}) reduces to a set of two, coupled equations
(one of which is merely algebraic) for the radial factors
$\varphi_{1,2}(p)$ of the independent components of
$\phi(\bm{p})$:$$2\,E(p)\,\varphi_2(p)+2\,Z^2(p^2)
\int\limits_0^\infty\frac{{\rm d}q\,q^2}{(2\pi)^2}\,V(p,q)\,
\varphi_2(q)=\widehat M\,\varphi_1(p)\ ,\qquad 2\,E(p)\,
\varphi_1(p)=\widehat M\,\varphi_2(p)\ ,\qquad q\equiv|\bm{q}|\
;$$herein, the configuration-space potential $V(r)$ aimed for
enters via the radial potential function $V(p,q)$:$$V(p,q)\equiv
\frac{8\pi}{p\,q}\int\limits_0^\infty{\rm d}r\sin(p\,r)\sin(q\,r)
\,V(r)\ ,\qquad r\equiv|\bm{x}|\ .$$At the Goldstone point
$\widehat M=0$ of the arising spectrum of mass eigenvalues
$\widehat M,$ the set decouples and~the algebraic equation implies
$\varphi_1(\bm{p})\equiv0.$ Thus, a single bound-state equation
determines $\phi(\bm{p})=\varphi_2(\bm{p})\,\gamma_5$:
$$E(p)\,\varphi_2(p)+Z^2(p^2)\int\limits_0^\infty\frac{{\rm
d}q\,q^2}{(2\pi)^2}\,V(p,q)\,\varphi_2(q)=0\ .$$From the
configuration-space representation of this relation, the sought
interquark potential follows~as$$V(r)=-\frac{\widetilde
T(r)}{\varphi_2(r)}\ ,$$with $\widetilde T(r)$ denoting the
Fourier transform of the effective kinetic term
$E(p)\,\varphi_2(p)/Z^2(p^2)$ for $Z(p^2)\ne0.$

\begin{figure}[b]\centering\begin{tabular}{cc}
\includegraphics[scale=1.40139,clip]{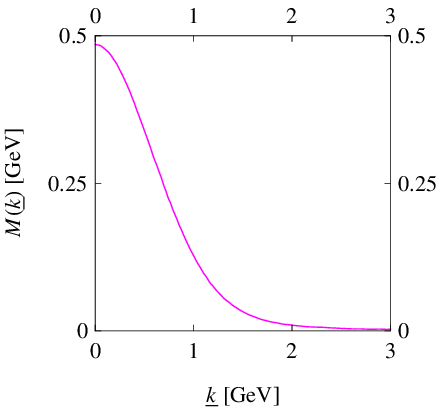}&
\includegraphics[scale=1.40139,clip]{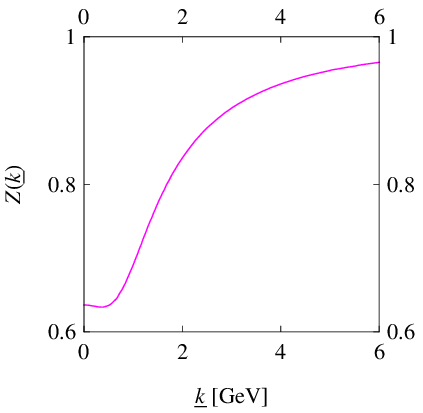}\\(a)&(b)
\end{tabular}\caption{Quark propagator functions: quark mass
function $M(\underline{k})$ (a) and wave-function renormalization
$Z(\underline{k})$~(b), plotted as functions of
$\underline{k}\equiv(\underline{k}^2)^{1/2}$ \cite{PM00}, from the
Dyson--Schwinger model for the quark propagator $S(\underline{k})$
of Ref.~\cite{PM99}.}\label{Fig:LSMZ}\end{figure}

\begin{figure}[t]\centering\includegraphics[scale=1.196952,clip]
{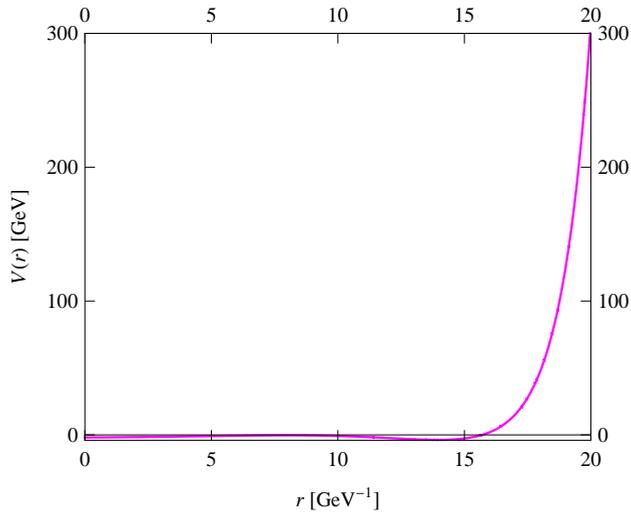}\caption{Configuration-space potential $V(r)$
from Fierz-symmetric effective interaction, by inverting the
Salpeter amplitude that emerges via Eq.~(\ref{Eq:W}) from the
quark-propagator solution of the Dyson--Schwinger model of
Ref.~\cite{PM99}.}\label{Fig:LSV}\end{figure}

Inverse problems of the type tackled here need, as informational
input, solutions of the bound-state equation one is concerned
with. In our case, this r\^ole is filled by $\varphi_2(\bm{p}),$
deduced from its ancestor $\Phi(p).$ We may harvest the latter
from Dyson--Schwinger analyses of the quark propagator in
Euclidean space (identified by \underline{underlining}) by
exploiting the fact \cite{PM97a} that, in the chiral limit, a
Ward--Takahashi identity connects full quark propagator and
flavour-nonsinglet pseudoscalar-meson Bethe--Salpeter
amplitude:\pagebreak \begin{equation}\Phi(\underline{k})\propto
\frac{Z(\underline{k}^2)\,M(\underline{k}^2)}
{\underline{k}^2+M^2(\underline{k}^2)}\,\underline{\gamma}_5+
\mbox{subleading contributions}\ .\label{Eq:W}\end{equation}
Seeking \emph{both\/} $M(p^2)$ \emph{and\/} $Z(p^2),$ we are led
to the Dyson--Schwinger solutions \cite{PM99} presented by
Ref.~\cite{PM00}.

With a well-defined starting point given in form of the quark
propagator functions of Fig.~\ref{Fig:LSMZ} at~hand, it is just a
matter of one's numerical skills to pin down the potential $V(r)$
at least in graphical form~\cite{WL16i}. Figure~\ref{Fig:LSV}
reveals the result: of all insights surely most essential is the
square-well shape of the potential.

\begin{figure}[h]\centering\begin{tabular}{cc}
\includegraphics[scale=1.234,clip]{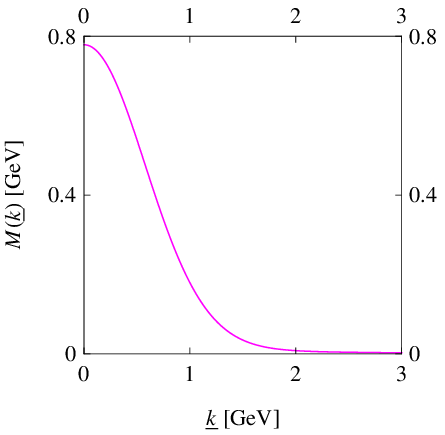}&
\includegraphics[scale=1.189,clip]{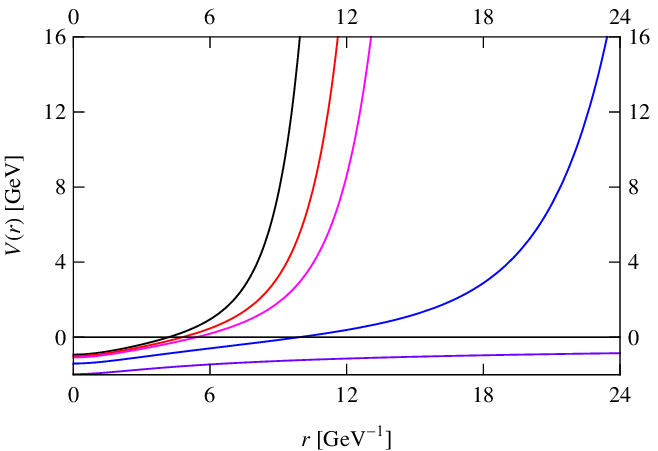}\\(a)&(b)
\end{tabular}\caption{(a) Mass function $M(\underline{k})$ derived
from the Dyson--Schwinger model of Ref.~\cite{PM97b} for the quark
propagator. (b) Associated configuration-space potential $V(r)$
from Fierz-symmetric kernel $K(\bm{p},\bm{q})$ for constituent
quark mass $m=0$ (black), $m=0.35\;\mbox{GeV}$
(\textcolor{red}{red}), $m=0.5\;\mbox{GeV}$
(\textcolor{magenta}{magenta}), $m=1\;\mbox{GeV}$
(\textcolor{blue}{blue}), and $m=1.69\;\mbox{GeV}$
(\textcolor{violet}{violet})~\cite{WL16n}.}\label{Fig:DSMV}
\end{figure}

Some limited insight about the behaviour of the interquark
potential can be gained by attributing to the propagator functions
$M(p^2)$ and $Z(p^2)$ entering into the three-dimensional
bound-state equation (\ref{Eq:F}) some prescribed constant values:
approximating the fermion propagators (\ref{Eq:P}) by their free
counterparts with effective constituent mass $m$ \cite{SE}
corresponds to considering the limit $M(p^2)\to m$ and $Z(p^2)\to
1$ of the approach of Sec.~\ref{Sec:L}. In this case, the $p_0$
integral of the static limit of Eq.~(\ref{Eq:BS}) may be given
anyway. From the solution for $M(\underline{k}^2)$ reported by
Ref.~\cite{PM97b} (Fig.~\ref{Fig:DSMV}(a)), we derive the
potentials $V(r)$~of Fig.~\ref{Fig:DSMV}(b), rising confiningly,
for $m$ below some crucial value, from a finite negative value at
$r=0$ to infinity~\cite{WL16n}.

\section{Configuration-space interquark potential: closing in on
analytic findings}\label{Sec:A}
\begin{figure}[b]\centering\includegraphics[scale=1.40962,clip]
{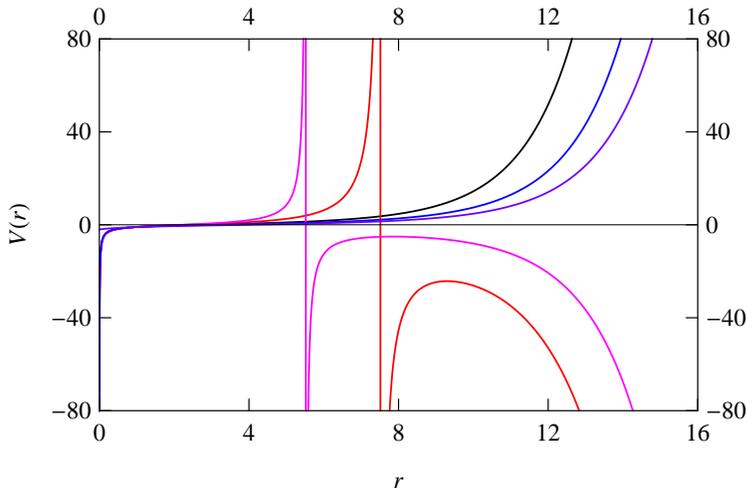}\caption{Configuration-space potential $V(r)$
from Fierz-symmetric effective interaction, for massless
constituent quarks and mixing parameter $\eta=0$ \cite{WL15}
(black), $\eta=1$ (\textcolor{red}{red}), $\eta=2$
(\textcolor{magenta}{magenta}), $\eta=-0.5$
(\textcolor{blue}{blue}), or
$\eta=-1$~(\textcolor{violet}{violet}).}\label{Fig:ARPm0}\end{figure}

\begin{figure}[t]\centering\includegraphics[scale=1.40962,clip]
{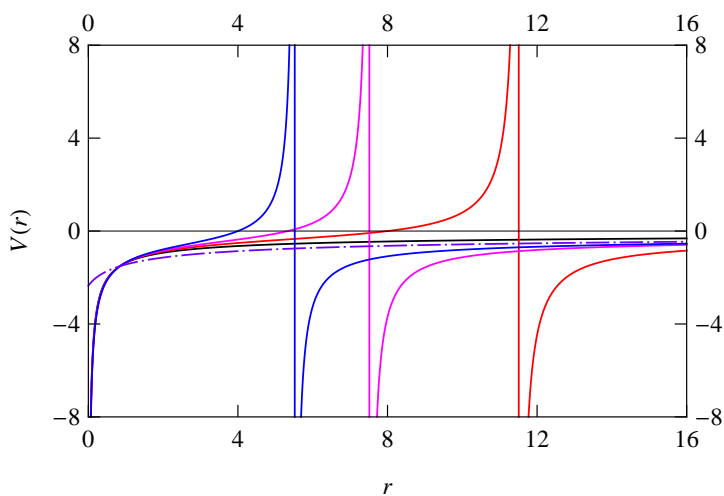}\caption{Configuration-space potential $V(r)$
from Fierz-symmetric effective interaction, for constituent
quarks~of mass $m=\mu$ and mixing value $\eta=0$ \cite{WL15}
(black), $\eta=0.5$ (\textcolor{red}{red}), $\eta=1$
(\textcolor{magenta}{magenta}), $\eta=2$ (\textcolor{blue}{blue}),
and $\eta=-1$~(\textcolor{violet}{violet}).}\label{Fig:ARPmu}
\end{figure}

Within the free-propagator
instantaneous Bethe--Salpeter formalism \cite{WL07}, we are able
to prove analytic statements about interquark potentials if we
place the focus on specific aspects of the quark propagator.
Dyson--Schwinger studies get that $M(\underline{k}^2)$ drops for
$\underline{k}^2\to\infty$ roughly like $1/\underline{k}^2$
\cite{PM97b}; axiomatic quantum field theory tells us that quark
confinement is ensured by the existence in $M(\underline{k}^2)$ of
an inflection point at $\underline{k}^2>0$ \cite{CDR08}. Both
aspects are captured by the ansatz \cite{WL15,WL16}, with mass
($\mu$) and mixing ($\eta$)~parameters,
$$\Phi(\underline{k})=\left[\frac{1}{(\underline{k}^2+\mu^2)^2}
+\frac{\eta\,\underline{k}^2}{(\underline{k}^2+\mu^2)^3}\right]
\underline{\gamma}_5\ ,\qquad\mu>0\ ,\qquad\eta\in\mathbb{R}\
.$$The analytic expressions of the inferred potentials $V(r)$ can
be found in Refs.~\cite{WL15,WL16}. The dependence of $V(r)$ on
$r$ (in units of $\mu$ and $\mu^{-1},$ respectively) is shown in
Figs.~\ref{Fig:ARPm0} (for $m=0$) and \ref{Fig:ARPmu} (for
$m=\mu$). Their most striking feature is, due to our ansatz, a
logarithmically softened Coulomb singularity at~the~origin.

\end{document}